\documentclass[usenatbib,usegraphicx]{mn2e}
\usepackage{times}

\title[Circular velocity profiles]{Circular velocity profiles of dark matter haloes}
\author[F. Stoehr]
{Felix Stoehr\\
Institut d'Astrophysique de Paris, 98bis Bd Arago, 75014 Paris, France\\
Max-Planck-Institut f\"ur Astrophysik, Karl-Schwarzschild-Str. 1, 85748 Garching, Germany\\
Email: stoehr@iap.fr}
\date{Accepted 2005 September 29. Received 2005 September 28; in original form from 2004 March 3}

\newcommand{\mnras}{MNRAS}
\newcommand{\nat}{Nature}
\newcommand{\apj}{ApJ}
\newcommand{\aj}{AJ}

\newcommand{\apjl}{ApJL}
\newcommand{\aap}{AAp}

\begin{document}
\maketitle

\begin{abstract}
We use a high-resolution simulation of a galaxy-sized dark matter halo, published simulated data as well as four cluster-sized haloes from Fukushige, Kawai \& Makino to study the inner halo structure in a $\Lambda$ cold dark matter cosmology. We find that the circular velocity curves are substantially better described by \citeauthor{Stoehr_et_al_02} (SWTS) profiles than by Navarro, Frenk \& White (NFW) or Moore et al. profiles. Our findings confirm that no asymptotic slope is reached and that the profiles are nearly universal, but not perfectly. The velocity profiles curve at a {\it constant} rate in $\log(r)$ over the full converged range in radii and the corresponding extrapolated density profiles reach a finite maximum density. We find, that the claim of a strong discrepancy between the observed inner slopes of the density profiles of low surface brightness galaxies (LSB) and their simulated counterparts on the grounds of currently available observations and simulations is unfounded. In addition, if the SWTS profile turns out to be a good description of the halo profile for the regions that cannot be probed with simulations of today, then even in these regions the agreeement between simulations and observations is very reasonable.

\end{abstract}

\begin{keywords} 
methods: N-body simulations -- galaxies: clusters: general -- galaxies: formation -- galaxies: haloes -- cosmology: theory -- dark matter.
\end{keywords}

\section{Introduction}
One of the greatest challenges to the $\Lambda$ cold dark matter (CDM) model favoured today is probably the apparent discrepancy between the very inner parts of observed dark matter (DM) density profiles and their counterparts in cosmological N-body simulations of structure formation. 

Numerical simulations seem to find nearly `universal' density profiles with cusps mostly being described by double-power-law fits with outer slope $\beta={\mbox d}\log\rho/{\mbox d}\log r\approx-3$ and inner slopes between -1 and -1.5. These profiles generally match the simulated curves reasonably well for scales of $r > 0.05 \ r_{200}$ where the virial radius $r_{200}$ is the radius of a sphere around the halo centre enclosing a density 200 times larger than the critical density (\citealt*{Navarro_Frenk_White_97}, NFW; \citealt*{Tormen_Bouchet_White_97}). The exact value of the inner slope, which had been constantly re-examined as computing power and thus mass resolution increased, is still a matter of debate (\citealt{Fukushige_Makino_97,Moore_Governato_Quinn_Stadel_Lake_98,Moore_et_al_99cold,Ghigna_et_al_00,Jing_Suto_00,Fukushige_Makino_01,Klypin_et_al_01,Jing_Suto_02,Fukushige_Makino_03,Power_et_al_03}). The best resolved haloes \citep*{Fukushige_Kawai_Makino_04} with up to $3 \times 10^7$ simulation particles within the virial radius showed slopes at the resolution limits of the simulations that were significantly shallower than -1.5.

Extrapolating the fits to smaller distances results in a strong discrepancy with the observed rotation curves of DM-dominated low surface brightness (LSB) galaxies. Their DM haloes are consistent with profiles having constant density cores \citep{Flores_Primack_94, Moore_94, Burkert_95, McGaugh_Blok_98, Firmani_et_al_01}.  Although the extent of this discrepancy seemed controversial \citep[e.g.][]{vandenBosch_Swaters_01}, the most recent studies \citep*{Blok_Gaugh_Rubin_01, deBlok_04} claim that simulations can not be reconciled with observations. Several observational effects like seeing, misalignment of the slit of the telescope, finite slit size, beam smearing due to the large beams of the radio telescopes or an offset between the dynamical and the optical centre certainly matter. However, they are too small to make the NFW profile with inner slope of -1 mimic observed rotation curves \citep{deBlok_04}. 

In this work we re-investigate the inner structure of DM haloes simulated at very high resolution and compare the results to published LSB galaxy data. In the next section we describe the simulation we have carried out and the simulated data we have used. We then give the results of the profile-fitting and discuss the implications. In the last section we summarise and discuss our conclusions.

\begin{figure*}
\centering \includegraphics[width=88mm]{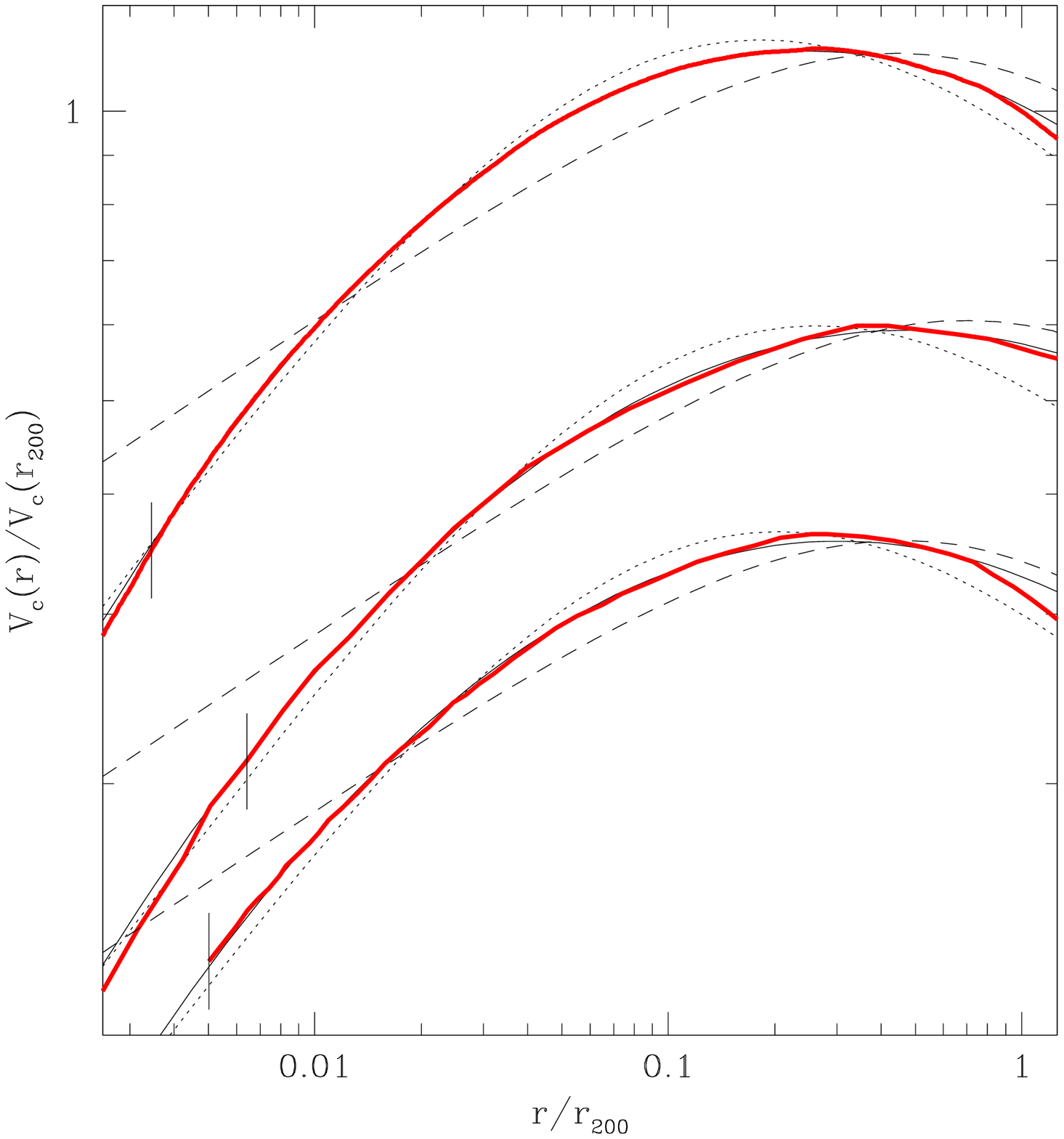}
\centering \includegraphics[width=88mm]{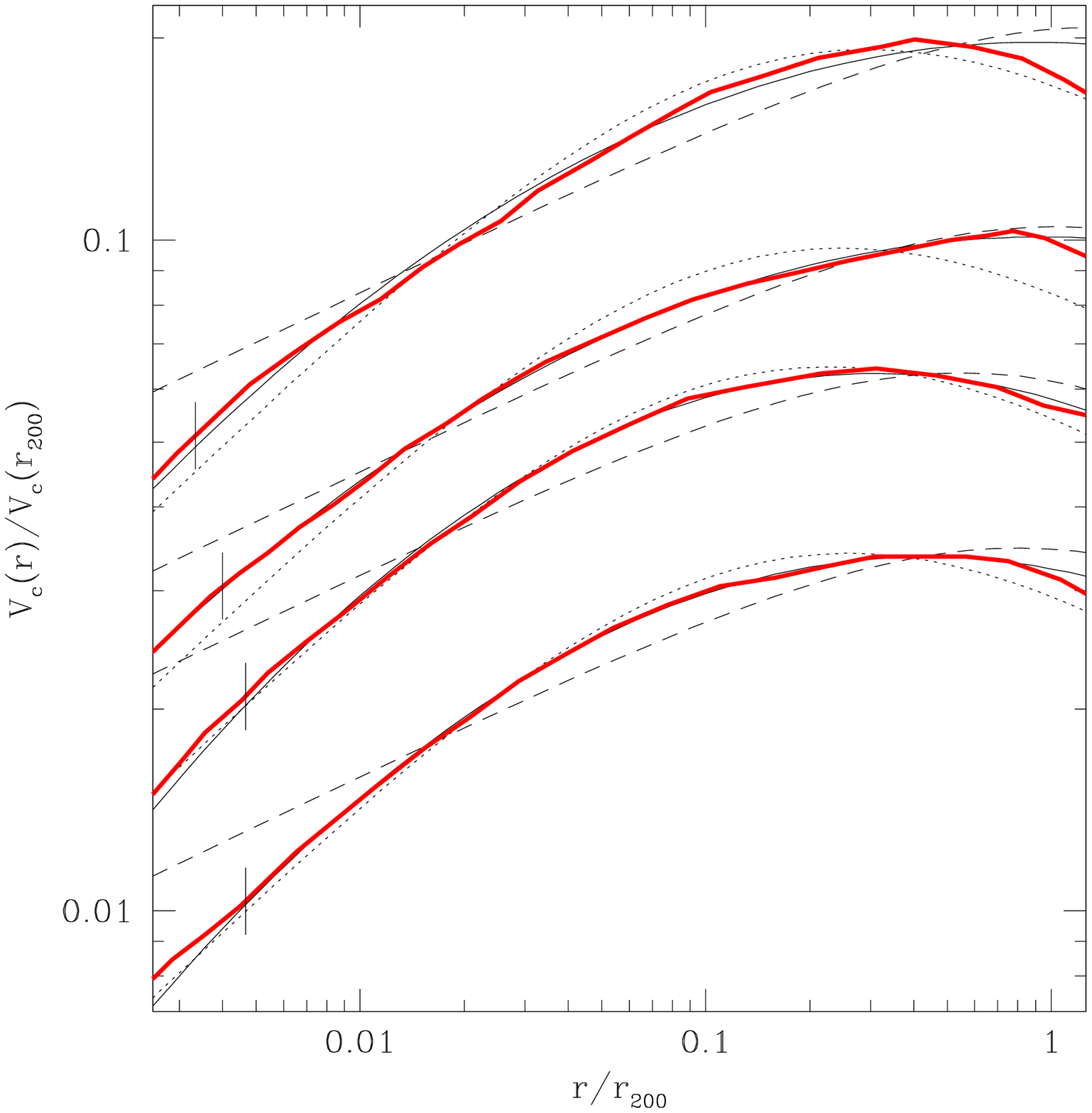}
\caption{{\it Left-hand panel:} circular velocity curves for GA3new, G3-256$^3$ \citep{Hayashi_et_al_04} and P256$^3$ \citep{Power_et_al_03} (thick solid, red) with over-plotted SWTS (thin solid), Moore (dashed) and NFW profiles (dotted). Vertical lines indicate the radius down to which the profiles are converged. {\it Right-hand panel:} circular velocity curves of the four cluster simulations carried out by \citet*{Fukushige_Kawai_Makino_04}. For clarity all curves have been shifted down by multiples of a quarter decade. The $a$-values of the seven SWTS profiles are       0.150, 0.128, 0.139, 0.102, 0.0934, 0.149 and 0.135. \label{figure:circularvelocity}}
\end{figure*}

\section{Simulations}
We reran the high-resolution simulation GA3n from \citet{Stoehr_et_al_03} with {\sc gadget-2}, a substantially improved, novel version of
{\sc gadget} \citep*{Springel_Yoshida_White_01} and termed it GA3new. This simulation is a two-level resimulation starting from a large cosmological volume (\citealt{Jenkins_et_al_01}; \citealt*{Yoshida_Sheth_Diaferio_01}) carried out in a flat $\Lambda$-dominated CDM cosmology with matter density $\Omega_m=0.3$, cosmological constant $\Omega_{\Lambda}=0.7$, expansion rate $H_0= 100 \, h \,$ km s$^{-1}$ Mpc$^{-1}$ with $h$~=~0.7, fluctuation amplitude $\sigma_8=0.9$ at $z=0$ and box side length $L=$~479~h$^{-1}$~Mpc. In the first-level resimulation, the mass resolution within a sphere of 52 h$^{-1}$ Mpc in diameter was increased by a factor of 411. 

In order to match the properties of the Milky Way, the halo candidate for the GA-simulation series selected from this first-level resimulation was carefully chosen to be isolated and to have had a very quiet merging history, i.e. the last major merging event happening before $z=1.5$ \citep{Stoehr_et_al_02}. In GA3new there are 11~562~566 particles within $r_{200}=217$ h$^{-1}$ kpc at $z=0$.

The new version of {\sc Gadget} combines the original tree-code with a particle-mesh (PM) method to compute the long-range forces. A 512$^3$-grid was used for the calculation of the corresponding fast Fourier transforms. We followed \citet{Power_et_al_03} for the choice of the numerical parameters but used timesteps half as large as required and in addition very conservative force parameters (i.e. ErrTolIntAccuracy=0.0125, ErrTolForceAcc=0.005). The softening length of $r_{eps}$~=~0.18 h$^{-1}$  kpc was kept constant in comoving coordinates. Due to the use of the PM method to compute the long range forces, the integration was extremely accurate also at high redshifts where the net forces on all particles nearly cancel out. 

We extracted from recently published data two additional circular velocity curves of galaxy-sized DM haloes with lower resolution. They were both produced using very similar techniques as that described above. The haloes G3-256$^3$ from \citet{Hayashi_et_al_04} and P256$^3$ from \citet{Power_et_al_03} have about $2.7 \times 10^6$ and $3.2 \times 10^6$ particles within their virial radii, respectively. 
Moreover, Toshiyuki Fukushige kindly made available the circular velocity curves of the four $\Lambda$CDM cluster simulations L1, L2, L3 and L4, presented in \citet*{Fukushige_Kawai_Makino_04}. These clusters contain $2.6 \times 10^7, 2.6 \times 10^7$, $7.2 \times 10^6$ and $7.8 \times 10^6$ particles inside their virial radii and were produced using GRAPE clusters and a GRAPE tree code \citep{Kawai_Makino_03}. 

\section{Results}
Fig. \ref{figure:circularvelocity} shows the circular velocity curves of all seven simulations offset by multiples of a quarter decade for clarity. Vertical lines indicate the radii $r_{conv}$ down to which the profiles have converged following the analysis of \citet{Power_et_al_03}. 

\begin{figure*}
\centering \includegraphics[width=88mm]{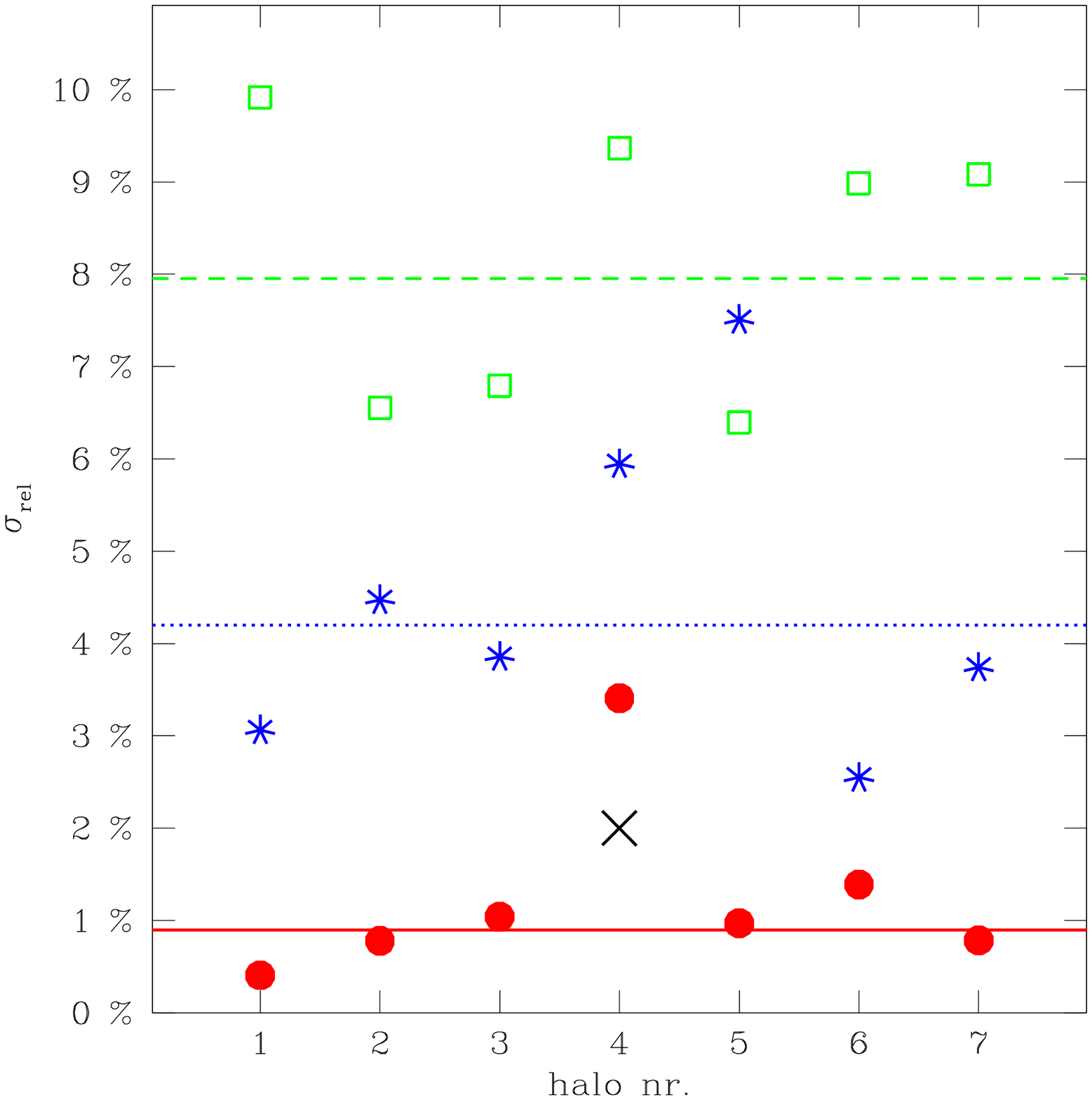}
\centering \includegraphics[width=88mm]{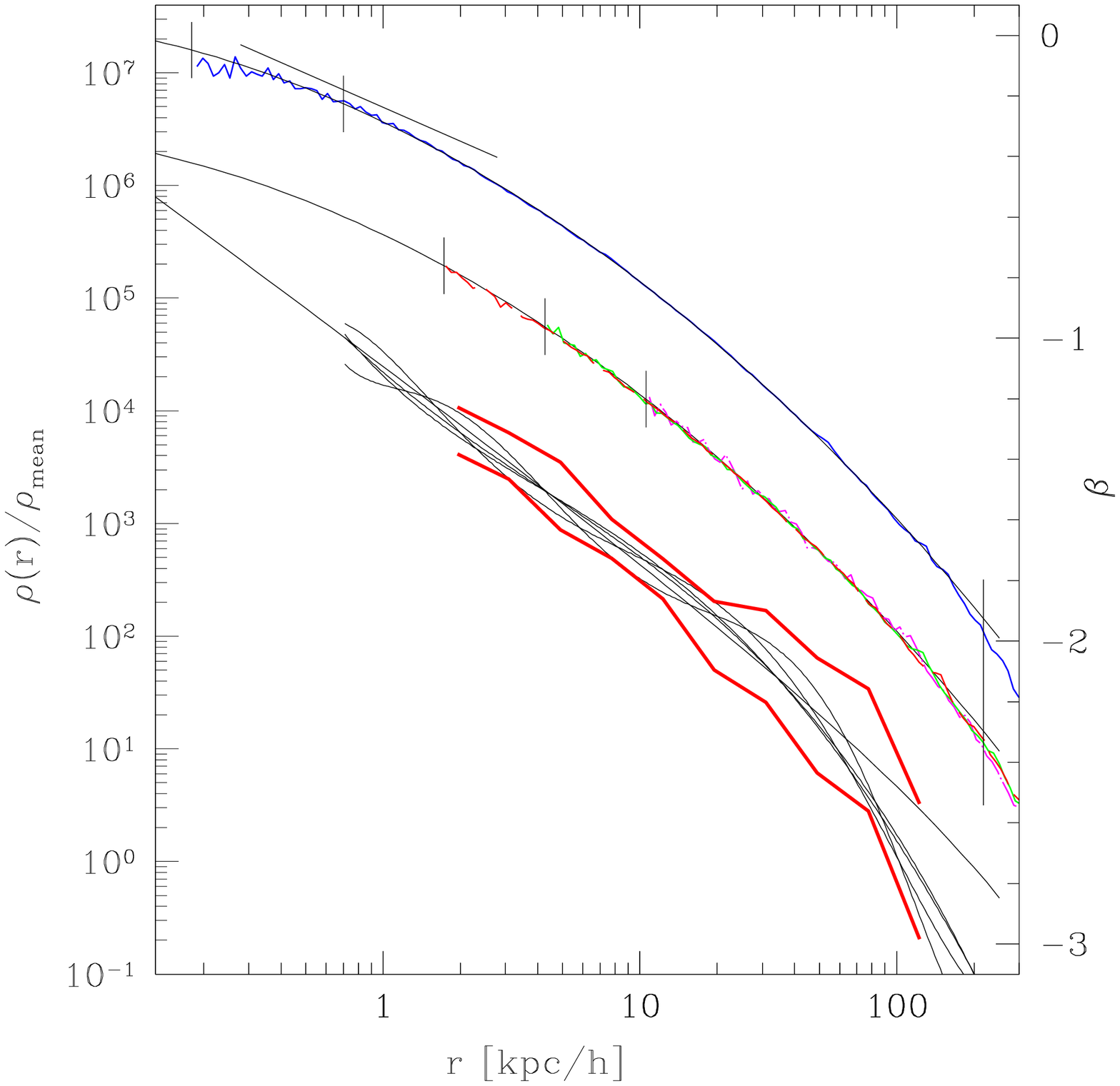}
\caption{{\it Left-hand panel:} standard relative deviations of the NFW (asterisks), Moore (squares) and SWTS (dots) profiles for all seven haloes. The lines indicate the average values (NFW: dotted, Moore: dashed, SWTS solid) excluding the values of the merging system L1 (halo number 4). {\it Right-hand panel:} density profiles of GA3new (solid), GA2n (long dashed), GA1n (dotted) and GA0n (solid) together with the SWTS density profile (smooth solid) corresponding to the best fit of GA3new in Fig. \ref{figure:circularvelocity} (i.e. the best-fitting combination of $a$, $r_{max}$ and $V_{max}$). The profiles of GA2n, GA1n and GA0n, i.e. the lower resolution runs of the GA-simulation series, are offset by one decade for clarity. The short diagonal line indicates the slope of -1. The slope $\beta$ of the SWTS density profile is shown by the lower diagonal line. Its values are given on the right-hand axis. The thick lines show the average slope envelope of a recent numerical study carried out by \citet{Reed_et_al_05} drawn down to its typical converged radius. Its radii were scaled down to match those of GA3new. The dotted lines show best-fit polynomials to the slope of GA3new. Vertical lines from left to right show the softening length $r_{eps}$, the converged radii $r_{conv}$ of GA3new, GA2n, GA1n and GA0n as well as the virial radius $r_{200}$ of the GA simulation series. \label{figure:qualityanddensity}}
\end{figure*}

For the comparison of simulated data with analytical profiles, we prefer circular velocity curves over density profiles for three reasons. Firstly, they typically span only about half a decade on the ordinate, which allows to identify by eye deviations from analytical fits in the {\it per cent} range. In addition, as they are a cumulative quantity, they suffer much less from Poisson noise. Finally, they are unique measures as no binning is necessary. 

A possible concern due to the cumulative nature of the circular velocity curves is that the profiles at radii somewhat larger than the converged radii could be affected by the not well-resolved inner parts of the haloes leading to underestimates of the inner slopes \citep{Kazantzidis_et_al_04}. We will address this concern below. 

The circular velocity curves in Fig. \ref{figure:circularvelocity} (thick solid lines) do not asymptotically approach a constant inner slope (i.e. do not have an ``asymptotic slope") but continue to curve all the way down to the converged radii. Such behaviour has been pointed out recently by several authors (\citealt*{Subramanian_Cen_Ostriker_00}; \citealt{Hayashi_et_al_04,Power_et_al_03,Tasitsiomi_et_al_04}; \citealt*{Hoeft_Muecket_Gottloeber_04}; \citealt{Navarro_et_al_04}; \citealt*{Diemand_Moore_Stadel_04}). Instead of a double-power-law profile, we therefore fit an analytical profile with continuously changing slope ${\mbox d}\log V_c/{\mbox d}\log r$ to the profile data. The simplest such profile is the \citefullauthor{Stoehr_et_al_02} (SWTS) profile \citep{Stoehr_et_al_02,Stoehr_et_al_03}, i.e. a parabola, where the slope changes at constant rate in $\log(r)$ (Fig. \ref{figure:circularvelocity}, thin solid lines):

\begin{equation}
\log\left(V_c/V_{max}\right) = - a \ \left[\log\left(r/r_{max}\right)\right]^2
\label{equation:circularvelocity}
\end{equation}

Details about the corresponding mass and density profiles are given in the Appendix. Note that the profile fits so well, that most of the time, the lines are covered by the simulated data over the full range in $r$ from the virial radius $r_{200}$ down to the converged radius $r_{conv}$.

All fits are done minimizing the standard relative deviation
\begin{equation}
\sigma_{rel} = \sqrt{\frac{\sum_{i=1}^n \left[\left(V_c(r_i)-V_{c,analytic}(r_i)\right)/ V_c(r_i)\right]^2}{n}}
\label{equation:meandeviation}
\end{equation}
measured at 100 points per decade in $r$. For GA3new, $\sigma_{rel}$ is only 0.4 per cent. 

For comparison we overplot also the best-fitting NFW \citep*{Navarro_Frenk_White_97} (dotted) and Moore profiles \citep{Moore_et_al_99} (dashed). These profiles, which have two not three adjustable parameters, show significant systematic deviations from the simulated data. For {\it all} of the seven haloes, the NFW profiles under-, over- and under-predict again the data at 1 percent of $r_{200}$, 10 percent of $r_{200}$ and $r_{200}$, respectively. For the Moore profiles the situation is inverted.

The left-hand panel of Fig. \ref{figure:qualityanddensity} shows for all seven haloes the standard relative deviations of the best-fitting NFW (asterisks), Moore (squares) and SWTS (dots) profiles.

Except for the most massive cluster L1, very good SWTS fits are possible. The cluster L1 is in a merging phase at $z=0$: the slope changes abruptly at 0.4 and at 5 per cent of $r_{200}$. In the hierarchical structure formation scenario, galaxy clusters are building up today and are thus, by construction, subject to recent merging events. This favours galaxy-sized DM haloes over cluster-sized haloes for the study of their inner structure. 

The average deviation between the data and the analytical fits is 4.2 per cent for the NFW profiles (dotted), 8.0 per cent for the Moore profiles (dashed), but only 0.89 per cent for the SWTS profiles (solid). We excluded here the merging cluster L1.

\begin{figure*}
\centering \includegraphics[width=88mm]{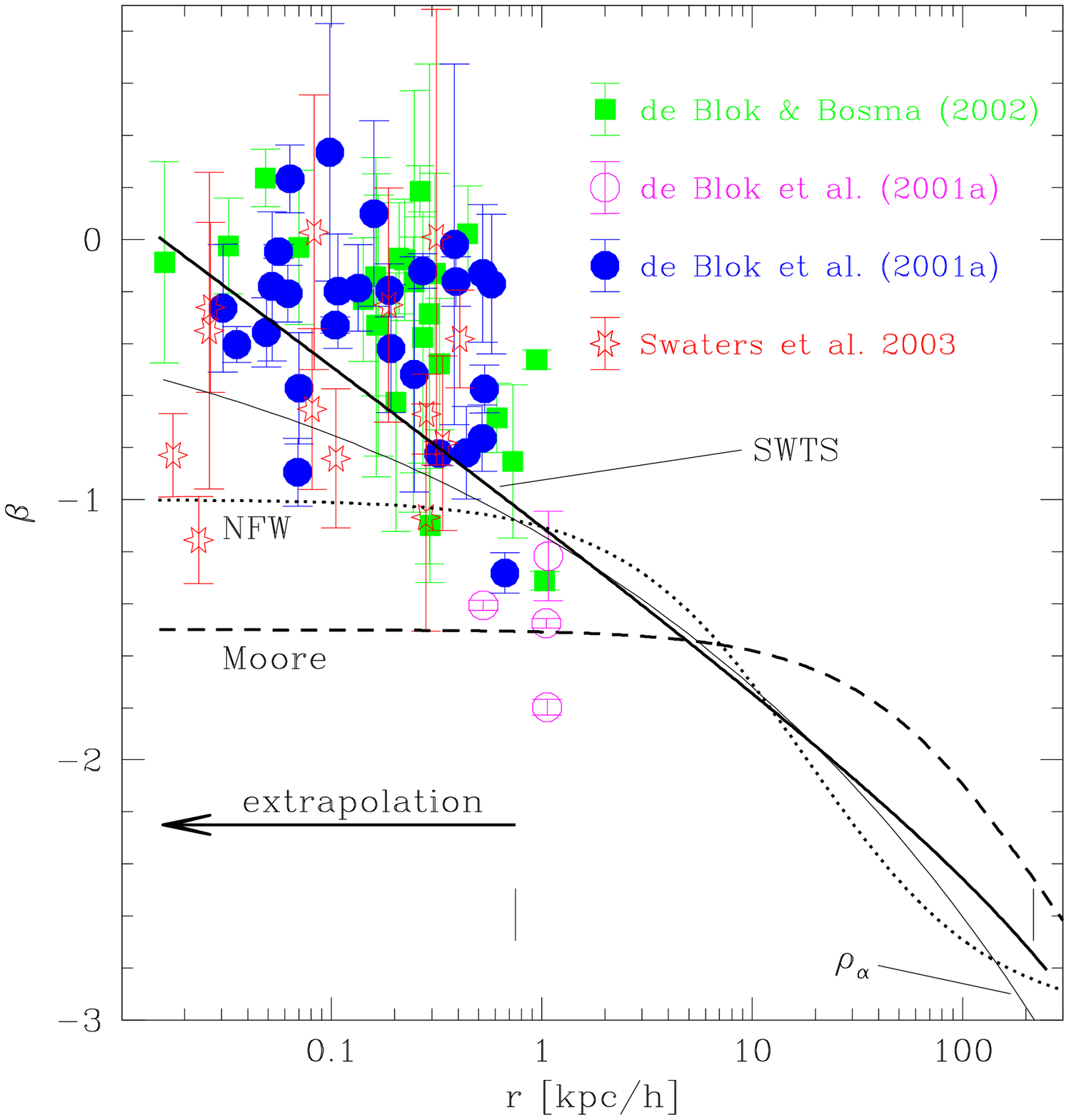}
\centering \includegraphics[width=88mm]{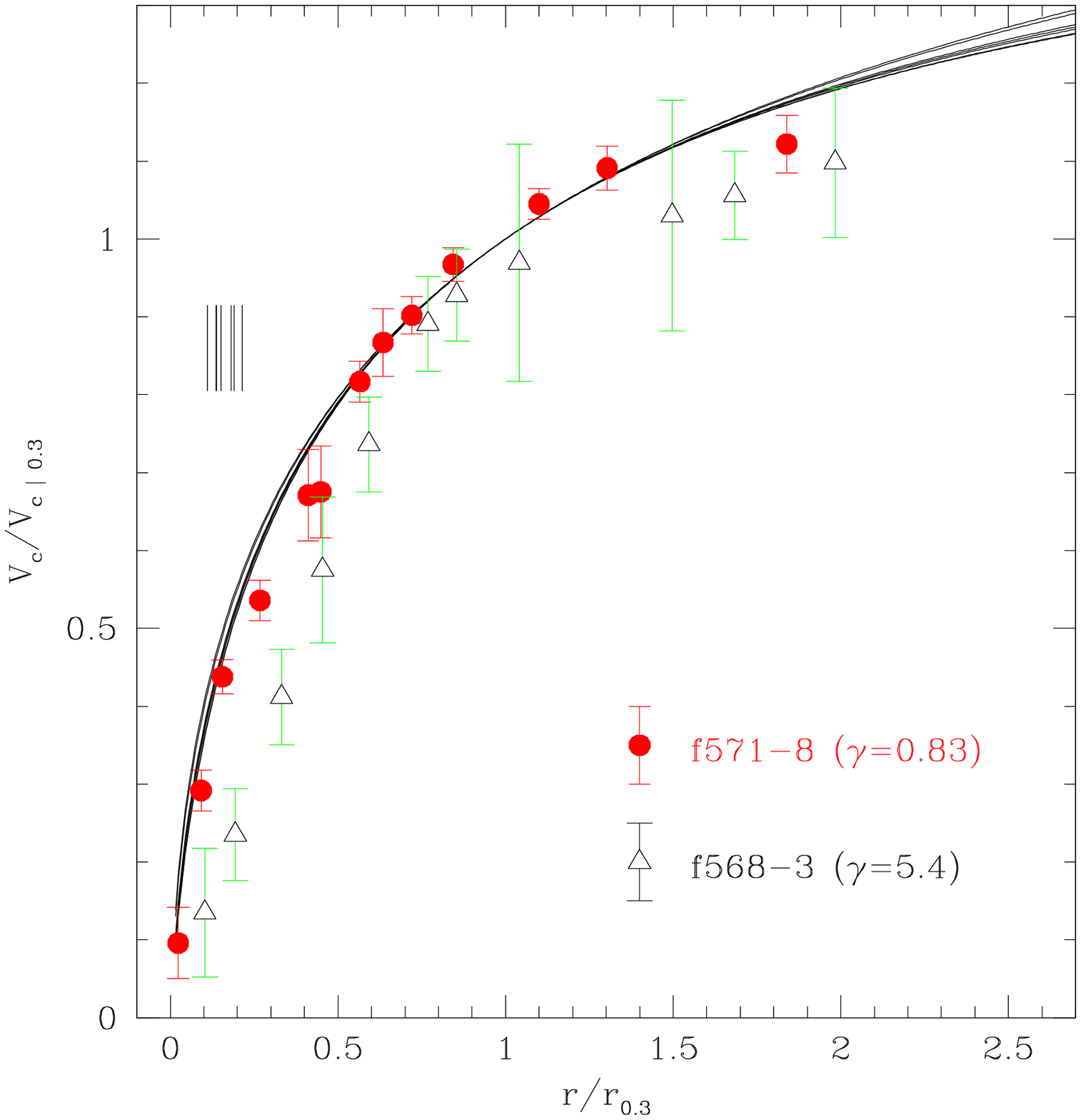}
\caption{{\it Left-hand panel:} inner density profile slopes of LSB galaxies taken from \citet{deBlok_04} who used data from \citet*{Blok_Gaugh_Rubin_01}, \citet{deBlok_Bosma_02} and \citet{Swaters_et_al_03}. We converted to h$^{-1}$ kpc with $h=0.7$. The logarithmic density profile slopes of the analytical circular velocity profile fits to GA3new of Fig. \ref{figure:circularvelocity} are shown as dotted (NFW), dashed (Moore) and solid (SWTS) lines. The profiles were extrapolated beyond $r_{conv}$ of GA3new (left-hand vertical line) as indicated by the arrow. The thin solid line shows the slope of the $\rho_{\alpha}$ profile recently proposed by \citet{Navarro_et_al_04}. {\it Right-hand panel:} two LSB galaxy rotation curves representing the bulk (dots) and the tail (triangles) of the shape distribution. The profiles are normalised to the point where the logarithmic slope equals 0.3 (see \citet{Hayashi_et_al_04} for details). Solid lines show all seven normalised SWTS profiles of Fig. \ref{figure:circularvelocity}. Solid vertical lines indicate the corresponding converged radii. \label{figure:lsb}}
\end{figure*}

As mentioned above, the unresolved very central parts of the DM distribution could lead to an underestimate of the circular velocity curve at small radii. In this case, the density profile would have a steeper slope at small radii than the density profile corresponding to the best SWTS fit (i.e. the best-fitting combination of $a$, $r_{max}$ and $V_{max}$) obtained from the circular velocity curve. 
It is clear from the right-hand panel of Fig. \ref{figure:qualityanddensity} that this is not the case. At the converged radius of GA3new, at $r_{conv}=0.7$ h$^{-1}$ kpc, the SWTS density profile (smooth solid line; equations \ref{equation:density}, \ref{equation:rcore} and \ref{equation:rhocore}) falls well on top of the density profile of GA3new (solid). 

The match is good even down to radii of about $0.5 \ r_{conv}$ corresponding to roughly 2 times the softening length $r_{eps}$. This makes sense, as for GA3new, $r_{conv}$ determined by the criterion of \citet{Power_et_al_03} is a conservative limit, as the integration and force accuracy of {\sc gadget} has improved substantially since the time their study was completed. 

We have offset the density profiles of the lower-resolution simulations of the GA convergence series (GA2n, GA1n and GA0n) by one magnitude for clarity (see \citet{Stoehr_et_al_02,Stoehr_et_al_03} for details). The smooth solid line shows again the SWTS density profile of GA3new.

It is also obvious from the right-hand panel of Fig. \ref{figure:qualityanddensity} that differences between the profile and the SWTS fit as small as those that can be spotted in the circular velocity plots are completely invisible in the density profile plot due to the Poisson noise in the radial bins and the compressed ordinate.

The slope $\beta$ of the SWTS density profile is nearly constantly changing in $\log(r)$ over more than 2 orders of magnitude as is indicated by the diagonal line with units given on the right. At the resolution limit of GA3new a slope of -1 is reached. We overplotted the envelope of the averaged density profile slopes as determined in a recent study by \cite{Reed_et_al_05} (thick solid lines). Their radial coordinate was scaled in order to match that of GA3new. 

The lower dotted lines show the best-fitting polynomials to the slope of the density profile of GA3new evaluated in 100 logarithmically spaced bins between the converged and the virial radii. The polynomials (here shown are the polynomials of orders 5 to 10) follow very well the SWTS prediction except for very large radii. There is substantial scatter in the functions due to the unavoidably noisy slope measurement. 

When extrapolating the SWTS density profile to very small radii, a {\it flat} density profile would be reached at $r_{flat}$ (see the Appendix). However, this is out of the reach of even the best resolved galaxy and cluster haloes ever carried out, GA3new and L1/L2. An isolated DM halo would need to be simulated with approximately 200 {\it billion} particles within $r_{200}$ in order to resolve $r_{flat}$. 

It seems reasonable to assume constant density for radii smaller than $r_{flat}$ where the density of the SWTS profile formally would decrease again. 

For radii larger than $r_{200}$, an extrapolation to the radius where the density equals the mean density of the Universe, seems appropriate. This radius is about 1 Mpc or approximately 5 $r_{200}$ for GA3new. Note however, that the SWTS profile in this region is just an upper bound as the density of the material outside of the accretion shock is lower than if it were virialised.

Our findings above have important consequences for the apparent discrepancy between simulated circular velocity profiles and observed LSB rotation curves. 

This is because this discrepancy in fact emerges not between the simulations and the observations themselves, but rather between the extrapolated analytical power-law  fits and the observations as has been also pointed out by \citet{Hayashi_et_al_04} and \citet{deBlok_04}. Extrapolation is necessary as even GA3new only is converged down to about 1 h$^{-1}$ kpc whereas observationally the innermost slopes of the LSB rotation curves are obtained on scales of 0.01 to 1 h$^{-1}$ kpc.

As we have shown above, the density profile continues to flatten towards smaller radii and no asymptotic slope is reached. This means that slopes at different radii are {\it necessarily} different. We argue here that much of the apparent discrepancy is a result of this comparison at different radii. 

The left-hand panel of Fig. \ref{figure:lsb} shows the logarithmic slopes of the density profiles of observed LSB galaxies as compiled by \citet{deBlok_04} who used data from \citet*{Blok_Gaugh_Rubin_01}, \citet{deBlok_Bosma_02} and \citet{Swaters_et_al_03}. We overplot the NFW, Moore and SWTS density profile slopes of GA3new. Vertical lines indicate $r_{conv}$ and $r_{200}$. 

We find that the overall agreement between the extrapolated SWTS profile slopes and the observed values is quite good. This is true even though neither a possible scatter in the $a$ value or the halo mass nor observational effects like slit misplacement or beam smearing have been taken into account. It has been shown that these observational uncertainties, all resulting slope in underestimates, probably are too small to account for the large discrepancy between observed slopes and NFW or Moore profiles \citep{deBlok_04}. However, they may be large enough to reduce the remaining discrepancy with the SWTS profiles. This discrepancy is much smaller: for about half of the LSB slopes, the GA3new profile is within the 1-$\sigma$ errors. 

The thin solid line in Fig. \ref{figure:lsb} shows the slope of the extrapolated $\rho_{\alpha}$ profile recently proposed by \citet{Navarro_et_al_04} that was fitted to the density profile of GA3new (third adjustable parameter $\alpha=0.18)$. This profile (just like the SWTS profile) does not reach an asymptotic slope and has been shown to be nearly identical to the SWTS profile in the converged regions of simulations \citep{Navarro_et_al_04}. For the the $\rho_{\alpha}$ profile, the agreement with the observed slopes is less good than that of the SWTS profile although it is still much closer to the observed values than it is the case for the NFW or Moore profiles.

Given the necessity for extrapolation as well as the observational uncertainities, it is clear from Fig. \ref{figure:lsb} that the claim that there is a strong discrepancy between the observed inner slopes of LSB galaxies and their simulated counterpartes does not hold. However, we show here in addition that, if the extrapolated SWTS profile is a good description of the actual DM halo profiles at small radii, then the agreement of observations and simulations is indeed astonishingly good.

As a word of caution we note that, although LSB galaxies are DM-dominated and analysis suggests that baryons do not have significant influence on the inner parts of the LSB DM profiles \citep{deBlok_04}, a confirmation with future simulations of LSB-type galaxies including realistic gas physics should be done.

Unfortunately, despite significant progress made by several authors \citep{Peebles_80, Hernquist_90, Syer_White_98, Nusser_Sheth_99,
Subramanian_et_al_00, Taylor_Navarro_01, Dekel_et_al_03}, it had not been possible yet to determine the shape of the dark matter profiles based on analytical arguments.

We now proceed from the innermost slopes to the overall profile shapes. The right-hand panel of Fig. \ref{figure:lsb} shows two LSB galaxy rotation curves \citep*{Blok_Gaugh_Rubin_01} representing the bulk and the tail of the LSB shape distribution (see \citet{Hayashi_et_al_04} for details). They are normalised to the point at which the logarithmic slope equals 0.3 \citep{Hayashi_et_al_04}. The `typical' profile, f571-8 (dots), is very well described by {\it all} seven normalised SWTS profiles of Fig. \ref{figure:circularvelocity} (solid). This is true even down to the innermost data points for which the profiles had to be extrapolated beyond their converged radii (upper vertical lines). 

The profiles of the simulated haloes fail to match the rotation curve of f568-3 from the tail of the profile shape distribution. The possible explanation, that LSB galaxies with profiles similar to that of f568-3 might be subject to very strong tidal fields that alter their inner structure, does not hold: the LSB galaxy samples contain isolated objects.

\section{Summary and Discussion}

We have reinvestigated the inner structure of DM haloes simulated in a $\Lambda$CDM cosmology using the best resolved galaxy and cluster haloes existing to date. The studied haloes were produced by different groups using three different simulation codes. We find that the concept of an `inner slope' of the circular velocity curves or density profiles is not appropriate: the profiles continue to curve without reaching an asymptotic slope in $\log(r)$. 

The simplest analytical velocity profile with such property, the profile with  constantly changing slope, is the SWTS profile. It describes the analysed DM haloes typically at a level of better than one per cent over the full range from the converged to the virial radius. This is substantially better than the values we obtain for NFW or Moore profiles.

Allowing for one more free parameter in the analytic profiles, the shape parameter $a$, is necessary in order to provide fits of high accuracy. It seems plausible that the small scatter in shape parameter values reflects the differences in non-sphericity of the DM distributions. We thus find that simulated dark matter profiles are nearly universal but not perfectly so. Our $a$ values range from 0.093 to 0.15 with a median value of $a_{med}$=0.135.

It is interesting that the same simple profile accurately describes both the profiles of isolated haloes as well as those of strongly perturbed substructure haloes \citep{Stoehr_et_al_02,Stoehr_et_al_03}.

There has been a claim of a strong apparent discrepancy between the innermost density profile slopes of observed LSB galaxies and their simulated counterparts. We find that this claim is unfounded. It relys on an extrapolation of the analytical profiles beyond the actual resolution limits of the numerical simulations. Doing so is particularily dangerous, as we have shown that NFW and Moore profiles, which were used to support the claim, show systematic deviations from the simulated profiles already in the numerically converged regions. 

Indeed, when extrapolating the SWTS profile (which fits the simulated profiles substantially better than the NFW or Moore profiles) instead, we find an astonishingly good agreement with the innermost profile slopes of observed LSB galaxies.

We show in addition that then the shape of a `typical' LSB rotation curve can be very well reproduced. At least some of the remaining substantial disagreement for the shapes and slopes of the LSB galaxies from the tail of the distribution may be explained with observational effects.

\noindent

\section*{Acknowledgments}
We thank Volker Springel for all the help and the effort he put into {\sc gadget} and {\sc gadget-2} as well as into the GA resimulation project. We also thank Toshiyuki Fukushige who kindly made his high-resolution circular velocity curves available for this work. We are indebted to Simon White for very useful advice and comments. We thank Avishai Dekel and Lucio Mayer for fruitful discussions and Nick Seymour for correcting the manuscript. This work was supported by the Research and Training Network `The physics of the Intergalactic Medium' set up by the European Community under contract HPRN-CT-2000-00126.

\appendix
\section[]{The SWTS profile}
The density profile corresponding to the SWTS circular velocity from equation (\ref{equation:circularvelocity}) and its logarithmic slope $\beta$ are
\begin{eqnarray}
\rho_{SWTS}(r) = \frac{V_{max}^2}{4 \pi G} \ 10^{-2 a \left[\log\left(\frac{r}{r_{max}}\right)\right]^2} \frac{1}{r^2} \times \nonumber \\
\times \ \left[1-4 \ a \ \log\left(\frac{r}{r_{max}}\right) \right]
\label{equation:density}
\end{eqnarray}
\begin{eqnarray}
\beta_{SWTS}(r) = -2 - \ 4 a \ \biggl\{ \log\left(\frac{r}{r_{max}}\right) + \nonumber \\ + \frac{1}{\ln(10) [1 - 4 a \log(r/r_{max})]} \biggr\}
\end{eqnarray}
for $r>r_{core}$ where $r_{core}$ is the radius at which the density profile becomes constant. This radius can be obtained by requiring that the mass interior to $r_{core}$ (at constant density) is the mass required to produce the rotation curve at $r_{core}$, i.e.
\begin{equation}
   \frac{4 \pi}{3} r_{core}^3 \rho_{SWTS}(r_{core}) = M_{core} = \frac{V(r_{core})^2 \ r_{core}}{G}.
\end{equation}
This gives
\begin{equation}
   r_{core}= r_{max} \ 10^{-\frac{1}{2 a}}.
   \label{equation:rcore}
\end{equation}
The maximal density of the profile is then
\begin{equation}
   \rho_{SWTS}(r_{core})  = \frac{3}{4 \pi G} \ \frac{V_{max}^2}{r_{max}^2}
  \ 10^{1/(2a)}.
  \label{equation:rhocore}
\end{equation}
This density is adopted for $r<r_{core}$. The corresponding $M(<r)$ and $V(r)$ are just ($r<r_{core}$)
\begin{eqnarray}
M(r<) = \frac{4 \pi}{3} \ r^3 \ \rho_{SWTS}(r_{core}) \\
  V(r)  =  r \ \ \sqrt{\frac{4 \pi}{3}  \ \ G \ \ \rho_{SWTS}(r_{core}) }
\end{eqnarray}
It is easy to cross-check, that the core radius is the radius at which the rotation curve has a slope of 1. The mass $M
(<r)$ for $r>r_{core}$ is 
\begin{equation}
   M(<r) = \frac{V^2 \ r}{G} = \frac{V^2_{max} \ 10^{-2a \left[\log(\frac{r}{r_{max}})\right]^2}\ r} {G}
\end{equation}
We note however, that the density profile is not smooth at $r_{core}$. The radius at which the density profile gets flat is a bit smaller than $r_{core}$ for typical values of $a$:
\begin{equation}
   r_{flat}= r_{max}\  {10^{-\frac{1}{8 a} \left[1+\sqrt{9 + 16 a / \ln(10)}\right]}}.
\end{equation}
The density at this point is
\begin{eqnarray}
   \rho_{SWTS}(r_{flat}) = \frac{1}{8 \pi G} \frac{V_{max}^2}{r_{max}^2}
 10^{\frac{3 \sqrt{9 + 16 a/\ln(10)} - 1 - 8 a / \ln(10)}{16 a}} \nonumber \\
 \times \ \left[3+\sqrt{9+16 a /\ln(10)}\right]
\end{eqnarray}
Using $r_{flat}$ instead of $r_{core}$, and thus obtaining a smooth density profile resulting in a smooth total rotation curve is not a bad approximation. For GA3new, the relative difference in mass integrated up to the maximum of the rotation curve is 4.0 $\times$ 10$^{-5}$.

For GA3new, $r_{flat}$, $r_{core}$, $r_{conv}$, $r_{max}$, $r_{200}$, $V_{max}$,  $V_{200}$ and $a$ are 0.0156 h$^{-1}$ kpc, 0.0216 h$^{-1}$ kpc, 0.701 h$^{-1}$ kpc, 54.58 h$^{-1}$ kpc, 217.27 h$^{-1}$ kpc, 251.22 km s$^{-1}$, 217.27 km s$^{-1}$ and 0.15, respectively.


\begin{thebibliography}{45}
\expandafter\ifx\csname natexlab\endcsname\relax\def\natexlab#1{#1}\fi

\bibitem[{Burkert}(1995)]{Burkert_95}
{Burkert} A., 1995, \apjl, 447, L25+

\bibitem[{de Blok}(2004)]{deBlok_04}
{de Blok} W.~J.~G., 2004, in S.~Ryder, D.J.~Pisano, M.~Walker, K.~Freeman, eds, Proc.  IAU Symp. 220, Dark Matter in Galaxies, Astron. Soc. Pac., San Francisco, p.69

\bibitem[{de Blok} \& {Bosma}(2002)]{deBlok_Bosma_02}
{de Blok} W.~J.~G., {Bosma} A., 2002, \aap, 385, 816

\bibitem[{de Blok} et~al.(2001){de Blok}, {McGaugh} \&
  {Rubin}]{Blok_Gaugh_Rubin_01}
{de Blok} W.~J.~G., {McGaugh} S.~S., {Rubin} V.~C., 2001, \aj, 122, 2396

\bibitem[{Dekel} et~al.(2003){Dekel}, {Arad}, {Devor} \&
  {Birnboim}]{Dekel_et_al_03}
{Dekel} A., {Arad} I., {Devor} J., {Birnboim} Y., 2003, \apj, 588, 680

\bibitem[{Diemand} et~al.(2004){Diemand}, {Moore} \&
  {Stadel}]{Diemand_Moore_Stadel_04}
{Diemand} J., {Moore} B., {Stadel} J., 2004, \mnras, 325, 1017

\bibitem[{Firmani} et~al.(2001){Firmani}, {D'Onghia}, {Chincarini}, {Hern{\'
  a}ndez} \& {Avila-Reese}]{Firmani_et_al_01}
{Firmani} C., {D'Onghia} E., {Chincarini} G., {Hern{\' a}ndez} X.,
  {Avila-Reese} V., 2001, \mnras, 321, 713

\bibitem[{Flores} \& {Primack}(1994)]{Flores_Primack_94}
{Flores} R.~A., {Primack} J.~R., 1994, \apjl, 427, L1

\bibitem[{Fukushige} et~al.(2004){Fukushige}, {Kawai} \&
  {Makino}]{Fukushige_Kawai_Makino_04}
{Fukushige} T., {Kawai} A., {Makino} J., 2004, \apj, 606, 625

\bibitem[{Fukushige} \& {Makino}(1997)]{Fukushige_Makino_97}
{Fukushige} T., {Makino} J., 1997, \apjl, 477, L9+

\bibitem[{Fukushige} \& {Makino}(2001)]{Fukushige_Makino_01}
{Fukushige} T., {Makino} J., 2001, \apj, 557, 533

\bibitem[{Fukushige} \& {Makino}(2003)]{Fukushige_Makino_03}
{Fukushige} T., {Makino} J., 2003, \apj, 588, 674

\bibitem[{Ghigna} et~al.(2000){Ghigna}, {Moore}, {Governato}, {Lake}, {Quinn}
  \& {Stadel}]{Ghigna_et_al_00}
{Ghigna} S., {Moore} B., {Governato} F., {Lake} G., {Quinn} T., {Stadel} J.,
  2000, \apj, 544, 616

\bibitem[{Hayashi} et~al.(2003){Hayashi}, {Navarro}, {Power}
  et~al.]{Hayashi_et_al_04}
{Hayashi} E., {Navarro} J.~F., {Power} C., et~al., 2004, \mnras, 355, 794

\bibitem[{Hernquist}(1990)]{Hernquist_90}
{Hernquist} L., 1990, \apj, 356, 359

\bibitem[{Hoeft} et~al.(2004){Hoeft}, {Muecket} \&
  {Gottloeber}]{Hoeft_Muecket_Gottloeber_04}
{Hoeft} M., {Muecket} J.~P., {Gottloeber} S., 2004, \apj, 602, 162

\bibitem[{Jenkins} et~al.(2001){Jenkins}, {Frenk}, {White}
  et~al.]{Jenkins_et_al_01}
{Jenkins} A., {Frenk} C.~S., {White} S.~D.~M., et~al., 2001, \mnras, 321, 372

\bibitem[{Jing} \& {Suto}(2000)]{Jing_Suto_00}
{Jing} Y.~P., {Suto} Y., 2000, \apjl, 529, L69

\bibitem[{Jing} \& {Suto}(2002)]{Jing_Suto_02}
{Jing} Y.~P., {Suto} Y., 2002, \apj, 574, 538

\bibitem[{Kawai} \& {Makino}(2003)]{Kawai_Makino_03}
{Kawai} A., {Makino} J., 2003, in Makino J. Hut P., eds, IAU Symp. 208, Astrophysical Supercomputing using Particle Simulations. Astron. Soc. Pac., San Francisco, p.305

\bibitem[{Kazantzidis} et~al.(2004){Kazantzidis}, {Mayer}, {Mastropietro},
  {Diemand}, {Stadel} \& {Moore}]{Kazantzidis_et_al_04}
{Kazantzidis} S., {Mayer} L., {Mastropietro} C., {Diemand} J., {Stadel} J.,
  {Moore} B., 2004, \apj, 608, 663

\bibitem[{Klypin} et~al.(2001){Klypin}, {Kravtsov}, {Bullock} \&
  {Primack}]{Klypin_et_al_01}
{Klypin} A., {Kravtsov} A.~V., {Bullock} J.~S., {Primack} J.~R., 2001, \apj,
  554, 903

\bibitem[{McGaugh} \& {de Blok}(1998)]{McGaugh_Blok_98}
{McGaugh} S.~S., {de Blok} W.~J.~G., 1998, \apj, 499, 41

\bibitem[{Moore}(1994)]{Moore_94}
{Moore} B., 1994, \nat, 370, 629

\bibitem[{Moore} et~al.(1999{\natexlab{a}}){Moore}, {Ghigna}, {Governato}
  et~al.]{Moore_et_al_99}
{Moore} B., {Ghigna} S., {Governato} F., et~al., 1999{\natexlab{a}}, \apjl,
  524, L19

\bibitem[{Moore} et~al.(1998){Moore}, {Governato}, {Quinn}, {Stadel} \&
  {Lake}]{Moore_Governato_Quinn_Stadel_Lake_98}
{Moore} B., {Governato} F., {Quinn} T., {Stadel} J., {Lake} G., 1998, \apjl,
  499, L5

\bibitem[{Moore} et~al.(1999{\natexlab{b}}){Moore}, {Quinn}, {Governato},
  {Stadel} \& {Lake}]{Moore_et_al_99cold}
{Moore} B., {Quinn} T., {Governato} F., {Stadel} J., {Lake} G.,
  1999{\natexlab{b}}, \mnras, 310, 1147

\bibitem[{Navarro} et~al.(1997){Navarro}, {Frenk} \&
  {White}]{Navarro_Frenk_White_97}
{Navarro} J.~F., {Frenk} C.~S., {White} S. D.~M., 1997, \apj, 490, 493+

\bibitem[{Navarro} et~al.(2004){Navarro}, {Hayashi}, {Power}
  et~al.]{Navarro_et_al_04}
{Navarro} J.~F., {Hayashi} E., {Power} C., et~al., 2004, \mnras, 349, 1039

\bibitem[{Nusser} \& {Sheth}(1999)]{Nusser_Sheth_99}
{Nusser} A., {Sheth} R.~K., 1999, \mnras, 303, 685

\bibitem[{Peebles}(1980)]{Peebles_80}
{Peebles} P. J.~E., 1980, The Large-Scale Structure of the Universe, Princeton

\bibitem[{Power} et~al.(2003){Power}, {Navarro}, {Jenkins}
  et~al.]{Power_et_al_03}
{Power} C., {Navarro} J.~F., {Jenkins} A., et~al., 2003, \mnras, 338, 14

\bibitem[{Reed} et~al.(2005){Reed}, {Governato}, {Verde} et~al.]{Reed_et_al_05}
{Reed} D., {Governato} F., {Verde} L., et~al., 2005, \mnras, 357, 82

\bibitem[Springel et~al.(2001)Springel, Yoshida \&
  White]{Springel_Yoshida_White_01}
Springel V., Yoshida N., White S. D.~M., 2001, New Astronomy, 6, 79

\bibitem[{Stoehr} et~al.(2003){Stoehr}, {White}, {Springel}, {Tormen} \&
  {Yoshida}]{Stoehr_et_al_03}
{Stoehr} F., {White} S.~D.~M., {Springel} V., {Tormen} G., {Yoshida} N., 2003,
  \mnras, 345, 1313

\bibitem[{Stoehr} et~al.(2002){Stoehr}, {White}, {Tormen} \&
  {Springel}]{Stoehr_et_al_02}
{Stoehr} F., {White} S.~D.~M., {Tormen} G., {Springel} V., 2002, \mnras, 335,
  L84

\bibitem[{Subramanian} et~al.(2000{\natexlab{a}}){Subramanian}, {Cen} \&
  {Ostriker}]{Subramanian_Cen_Ostriker_00}
{Subramanian} K., {Cen} R., {Ostriker} J.~P., 2000{\natexlab{a}}, \apj, 538,
  528

\bibitem[{Subramanian} et~al.(2000{\natexlab{b}}){Subramanian}, {Cen} \&
  {Ostriker}]{Subramanian_et_al_00}
{Subramanian} K., {Cen} R., {Ostriker} J.~P., 2000{\natexlab{b}}, \apj, 538,
  528

\bibitem[{Swaters} et~al.(2003){Swaters}, {Madore}, {van den Bosch} \&
  {Balcells}]{Swaters_et_al_03}
{Swaters} R.~A., {Madore} B.~F., {van den Bosch} F.~C., {Balcells} M., 2003,
  \apj, 583, 732

\bibitem[{Syer} \& {White}(1998)]{Syer_White_98}
{Syer} D., {White} S. D.~M., 1998, \mnras, 293, 337+

\bibitem[{Tasitsiomi} et~al.(2004){Tasitsiomi}, {Kravtsov}, {Gottloeber} \&
  {Klypin}]{Tasitsiomi_et_al_04}
{Tasitsiomi} A., {Kravtsov} A.~V., {Gottloeber} S., {Klypin} A.~A., 2004,
 \apj, 607, 125

\bibitem[{Taylor} \& {Navarro}(2001)]{Taylor_Navarro_01}
{Taylor} J.~E., {Navarro} J.~F., 2001, \apj, 563, 483

\bibitem[{Tormen} et~al.(1997){Tormen}, {Bouchet} \&
  {White}]{Tormen_Bouchet_White_97}
{Tormen} G., {Bouchet} F.~R., {White} S. D.~M., 1997, \mnras, 286, 865

\bibitem[{van den Bosch} \& {Swaters}(2001)]{vandenBosch_Swaters_01}
{van den Bosch} F.~C., {Swaters} R.~A., 2001, \mnras, 325, 1017

\bibitem[{Yoshida} et~al.(2001){Yoshida}, {Sheth} \&
  {Diaferio}]{Yoshida_Sheth_Diaferio_01}
{Yoshida} N., {Sheth} R.~K., {Diaferio} A., 2001, \mnras, 328, 669

\end{thebibliography}
\end{document}